# 3D atomic structure from a single XFEL pulse

G. Bortel[1*], M. Tegze[1], M. Sikorski[2], R. Bean[2], J. Bielecki[2], C. Kim[2], J. Koliyadu[2], F. Koua[2], M. Ramilli[2], A. Round[2], T. Sato[2], D. Zabelskii[2], G. Faigel[1*]

[1] Wigner Research Centre for Physics, Institute for Solid State Physics and Optics, P.O.B. 49, 1525 Budapest, Hungary

[2] European XFEL GmbH, Holzkoppel 4, 22869 Schenefeld, Germany

\* Corresponding authors: bortel.gabor@wigner.hu, gf@szfki.hu

Abstract

**X-ray Free Electron Lasers (XFEL) are the most advanced pulsed x-ray sources. Their extraordinary pulse parameters promise unique applications. Indeed, several new methods have been developed at XFEL-s. However, no methods are known, which would allow *ab initio* atomic level structure determination using only a single XFEL pulse. Here, we present experimental results, demonstrating the determination of the 3D atomic structure from data obtained during a single 25 fs XFEL pulse. Parallel measurement of hundreds of Bragg reflections was done by collecting Kossel line patterns of GaAs and GaP. With these measurements, we reached the ultimate temporal limit of the x-ray structure solution possible today. These measurements open the way for studying non-repeatable fast processes and structural transformations in crystals for example measuring the atomic structure of matter at extremely non-ambient conditions or transient structures formed in irreversible physical, chemical, or biological processes. It would also facilitate time resolved pump-probe structural studies making them significantly shorter than traditional serial crystallography.**

Introduction

XFEL sources were built to widen our knowledge of the matter at extremes. These sources promise the solution of structures of extremely small samples, even single molecules or small atom clusters[1], and the measurement of structural changes at extremely short timescales like 1-100 fs. In the last ten years, we have seen significant improvements in single particle imaging (SPI)[2,3]. However, the original expectations are not yet met. The resolution of these measurements is far from atomic. Another type of experiment for structural studies at XFEL-s is serial femtosecond crystallography (SFX)[4,5]. It is based on the same concepts as SPI: 1.

measure before destruction and 2. serial measurements on thousands or millions of similar samples and combine the measured patterns to a single 3D intensity distribution. SFX is capable of producing datasets good for *ab initio* structure solution at the atomic level and it is widely used in biology, physics and chemistry. SFX works better than SPI because it uses small crystallites with periodic structures and not an arbitrary arrangement of atoms like SPI. So far, we discussed the study of static structures. If we would like to go further and measure structures changing in very short time scales, we can use SPI and SFX but only in the pump-probe mode. This restricts these studies to processes, which can be repeated millions of times exactly the same way. Therefore, we are mostly limited to laser pulse-driven processes. The *ab initio* solution of structures at extremely non-ambient conditions like very high magnetic field or very high pressure where one cannot repeat the same experiment exactly the same way is unreachable today. To solve the structure of matter in these cases one needs a method, which provides a dataset complete enough for *ab initio* structure solution from a single XFEL pulse. In this paper, we report a demonstration experiment that gives us the 3D atomic structure of crystalline materials from data taken during a single 25 fs XFEL pulse. To achieve this, we measured Kossel line patterns, collecting many Bragg reflections simultaneously. Due to interference with internal sources, in favorable cases, this method can give us not only the amplitudes of the structure factors but also their phases. This further simplifies the structure solution. In these demonstration experiments, we collected Kossel patterns of good-quality crystals of GaP and GaAs and could determine both the amplitudes and the phases of the structure factors. Therefore, the electron density was obtained with a direct Fourier synthesis. Although in these demonstration experiments, we used large, almost perfect single crystals, such measurements could be extended to non-perfect and small crystals. In these cases, we might lose the phase information, but we can turn back to the traditional evaluation methods of single crystal diffraction, where only the amplitude of the structure factor is used for the structure solution. Further developments in the experimental setup, especially on the detector side, would allow atomic resolution holography measurement[6,7,8]. This requires a very similar setup as the Kossel line measurement, and it would extend structural studies to systems with orientation order only.

**Kossel line pattern**

Kossel lines are formed when atoms of a single crystal emit x-ray photons, and these are scattered by the crystal itself. This process is depicted schematically in Figure 1 and a Supplemental Movie. It is clear from the geometry and Bragg's law that the regions of modified intensity are conical sections on a flat 2D pixelated detector. Though the formation of Kossel lines was experimentally shown by Kossel, Loeck and Voges in 1935[9] and a theoretical description was given by Laue in the same year[10], the method was not widely used in structural studies. The reason for this is that experimentally it is much more involved than traditional x-ray diffraction, it is limited to samples containing heavy atoms capable of emitting x-rays and the theoretical treatment of these patterns is complicated. Further, in normal circumstances (ambient conditions and long measuring times) it does not give more information than single-crystal x-ray diffraction. However, with the introduction of very intense pulsed sources (XFEL-s) and advanced 2D pixelated detectors, the simultaneous measurement of many Bragg reflections without the need of rotating the crystal, which the Kossel technique provides, could lead to unique applications. One could collect a full diffraction data set in a very short time, in extreme, during a single XFEL pulse. This facilitates the measurement of fast processes in single crystals. Since data collection is within a single pulse, we can think of *ab initio* determination of the changing structure during non-repeatable processes. Presently, this is not possible by any other method. Further, having good quality crystals the fine structure of Kossel lines can provide the phases of the structure factors, eliminating the crystallographic phase ambiguity problem present in traditional crystallography. A theoretical description of the fine structure of Kossel lines for the x-ray Bragg case and in infinitely thick crystals has been developed by Hutton, Trammell and Hannon[11] within the framework of the dynamical theory of x-ray diffraction. However, for our purposes, this description is not enough, because we have finite thickness and many Laue case cones. Therefore, we turned to earlier works developed for nuclear sources and scatterers[12,13,14]. The theory worked out in these papers included the more general case of polarization mixing by the scattering medium and higher multipolarity sources. Since in our measurements, the sources are isotropic E1 emitters and the scattering is by electrons (*i.e.* electric dipole), we used a simplified form of the expressions given in these papers. For simplicity, we omitted polarization, which has negligible effects. Further, we applied the formulas for crystal slabs with finite thicknesses. We resorted to numerical solutions both in the Laue and in the Bragg case. With these modifications, the fine structure of the Kossel lines could be analyzed; the amplitudes and phases for the measured reflections were obtained. A more detailed description of the Kossel lines and the formulas used in the evaluation process are given in the Supplementary Information and their derivation in references[12,13,14].

**Experiment**

We have developed the experimental procedure for x-ray holography for many years at synchrotron sources and could reach measuring times in the range of 1 second for a statistically meaningful pattern[8]. Kossel line pattern measurements require a very similar setup as inside source holography[8,15,16,17], except the spatial resolution of the 2D detector used for parallel detection of the fluorescent intensity forming the Kossel lines or the hologram. The short measuring time of holograms and Kossel lines at synchrotrons[18] prompted us to try the measurement of Kossel lines at XFEL sources. We realized the possibility of collecting a pattern during a single pulse. Starting from intensity considerations only, this conclusion seems trivial, since in the probe beam at a synchrotron we have about the same number of photons during 1 second as we have in a single XFEL pulse. However, the collection of all these photons in the very short time of an XFEL pulse with the precision necessary for distinguishing the lines from the background is not straightforward. The reason is that the detectors used at synchrotrons are counting detectors, while their XFEL counterparts are charge-integrating detectors. We have already experienced at synchrotrons that the detector is the weak point in the measurement[18]. Since at XFEL-s, the detector problem is even more pronounced, we tried to optimize all other experimental conditions for this demonstration experiment. First, we used samples (GaAs, GaP) from which we already collected good Kossel patterns or expected good-quality patterns. This allows us to check and strengthen the validity of XFEL measurements. Second, we choose the incident energy (10.5 keV) to excite only one element of the sample (Ga), increasing this way the signal to background ratio. Third, the sample was placed in He atmosphere in order to decrease air scattering. Unfortunately, we could not optimize two more parameters, the sample thickness and the experimental geometry. In the experiment, we used 100 micron thick samples in transmission geometry (Figure 2 top left). None of the reflection geometries (Figure 2 top right) were available due to technical limitations and the sample was thicker than the optimal 20-30 microns. Higher intensities obtained from thinner samples or in reflection geometry would cause saturation and malfunction of the detector. The implemented forward transmission arrangement (Figure 2 bottom photo) yielded about 20/120 and 200/800 fluorescent photons/pixel/pulse at the edges/center of the 4M Jungfrau detector[19] placed at 120 mm from the sample for GaAs and GaP, respectively. Due to the mismatch between the detector speed and the intra-train pulse repetition rate only a single pulse was used from each train, namely data was acquired at 10 Hz. The spot size on the sample was set by compound refractive lenses to ~25 μm diameter. In one shot we had about 1 mJ total energy. Since pulses have different total energies because of the stochastic nature of the spontaneous emission, we took several shots, every shot at a new place of the sample to avoid the effect of radiation damage. With these beam parameters we do not expect distortion of the Kossel lines caused by radiation damage. (The radiation damage and possible nonlinear effects are discussed in more detail in the Supplementary Information).The sample motion was controlled by a fast x-y scanner, while the sample surface was checked by an optical microscope. Good shots were selected by visual inspection and statistical analysis of the recorded detector images later in the evaluation process.

**Results**

We have measured Kossel line patterns of GaAs and GaP single crystal samples using single 25 fs long XFEL x-ray pulses. By careful analysis of the Kossel line patterns, we could solve the structure of these samples *ab initio*. We arrived at the atomic structure in several steps: (i) background removal, (ii) Kossel line indexing, (iii) geometry refinement, (iv) profile extraction, (v) line profile fitting, (vi) Fourier transform of the complex structure factors. Details of these steps are described in the Supplementary Information. After background correction of the raw data, we identified and indexed the lines[20]. In Figure 3, left panel the indexed Kossel lines of GaAs produced by the Ga K$\alpha_{1,2}$ fluorescent photons are shown on a sphere together with the detector projected to this surface. Although lines caused by the Kβ radiation are also visible on the image, we do not show them here for two reasons: (i) they are much weaker, and therefore their line shape could not be extracted with as good quality as that of the K$\alpha_{1,2}$ lines (ii) we have not used them in obtaining the atomic structures. On Figure 3, right panel an enlarged part of the measured pattern is shown without the theoretically calculated lines. Kossel lines can be clearly seen. Precise refinement of the Kossel line geometry, which is a prerequisite of the line-profile extraction yielded cubic lattice symmetry with lattice parameters of 5.655 Å and 5.453 Å for GaAs and GaP, respectively, in good agreement with literature values[21,22]. About 100 extracted line profiles had clearly identifiable, statistically meaningful fine structure (see a detailed inventory in the Supplementary Information). These profiles were fitted with a theoretical Kossel line profile based on the dynamical theory of diffraction and amplitudes and phase angles of the reflections were obtained (see details in the Supplementary Information). In Figure 4 we show four typical line profiles of GaAs, two Bragg and two Laue cases for low, medium and high index reflections. Similarly, in Figure 5 four line profiles for GaP are depicted. In the Fourier synthesis, we included all ~100 amplitudes and phases from the fitted profiles extending up to a cubic index of 8. All the other structure factors were taken as unknown and included with zero amplitudes. The assumed Friedel symmetry ensured the obtained electron density to be real. The forward scattering amplitude was taken as 184 for GaP and 256 for GaAs to define the zero level of the density.

Since we measured many shots on the same sample, we could double-check our results. We choose three shots from each sample and we analyzed them independently. The exact number of fitted line profiles depended on the individual shot, because of the different statistics. The number of fitted profiles fluctuated between 85 and 120. In the reconstructed real space structure, the fluctuation in the number of used lines (the number of structure factors) did not cause any significant deviation. Figure 6 and the corresponding Supplementary Movie show the reconstructed 3D real space structure of GaAs and GaP. Their structures are almost identical, except for the lattice spacing and the atom types. In the case of GaAs, Ga and As atoms could not be distinguished within the experimental error. This is not surprising since Ga has 31, while As has 33 electrons to scatter. Therefore, their scattering strengths are very similar. However, by improving the experimental conditions (using reflection geometry and detectors capable of handling more photons on the full surface of the detector), statistics will be much better. In future experiments, this will allow for extracting more precise structure factors leading to the possibility of distinguishing elements having atomic numbers close to each other. In the case of GaP, the two atom types have significantly different numbers of electrons, which reflects in the real space reconstruction, and the Ga and P atoms can be easily distinguished. To conclude this paragraph, we would like to point out that our demonstration measurement not only provides an opportunity to determine the atomic structure of phases formed during very fast, non-repeatable processes, but also lays the foundation for holographic measurements with atomic resolution[7]. It will widen the family of materials for single pulse structure determination to systems not having translation periodicity, only orientation order. So far, we mentioned those applications, which are unique to this method. However, there is a possibility to use our single pulse imaging technique in pump-probe experiments at XFEL-s. As it is clear from the description of the experiment, we could pump the sample by a laser beam before the X-ray pulse hits it, the same way as it is done in time resolved SFX experiments. However, in our case we do not have to measure tens of thousands of images of new samples but one pattern only (or a few taking into account the stochastic nature of the XFEL pulses and the less than 100% hit rate). This would significantly shorten the total measuring time, saving this way the very expensive XFEL beamtime.

**Conclusions**

The very intense and short XFEL pulses provide unique possibilities to deepen our understanding of various forms of matter. However, full exploitation of these possibilities requires special methods. We developed an experimental setup and the proper evaluation tools, which use a single very short and intense XFEL pulse for *ab initio* structure determination. According to our knowledge, no other technique is capable of this. We demonstrated on GaAs and GaP single crystals that it is possible to solve their structure from data taken during a single 25 fs XFEL pulse. We recorded Kossel line patterns of these materials and used these for structure solution. Instead of a conventional crystallographic structure solution, we also demonstrated that the phases of structure factors can be extracted from the profile of Kossel lines, allowing direct Fourier synthesis of the electron density, and avoiding the phase ambiguity present in crystallography. This method opens new avenues in atomic-level structural studies. It might allow the study of crystalline phases formed during non-repeatable processes, facilitating measurements at extremely non-ambient conditions. Further, it gives the bases to atomic resolution x-ray holographic measurements widening the types of materials suitable for single pulse structure determination. It could also make time resolved pump-probe experiments on single crystals with sizes ~10 μm or larger at XFEL sources significantly shorter.

**Supplementary Information**

Supplementary Information is available for this paper.

**Data Availability**

The data that supports the plots within the paper and other findings of this study are available from the corresponding author on reasonable request. The complete experimental data becomes publicly available at the European XFEL GmbH after an embargo period (SPB beamline, 202202/p003051 proposal).

**Code Availability**

The procedures of the evaluation are described in the Supplementary information and our previous publications. The custom programming code used in the evaluation process is in development and unsuitable for public release. However, the programs used in this study are available from the corresponding author on reasonable request.

**Acknowledgements**

We would like to thank B. Pécz and J. Volk for carrying out the thinning of GaAs and GaP wafers. Parts of this research were carried out at the European X-ray Free Electron Laser facility in Hamburg.

**Author contributions**

All authors took part in the measurements at XFEL. In addition, G.F. came up with the idea of doing this measurement. He also did the theoretical work and the programming on the fine structure of the Kossel lines. He took part largely in the writing of the manuscript. G.B. developed the methods and software for online analysis, processing measured patterns, which allowed the removal of background, indexing the Kossel lines, refining the patterns and extracting the profile curves. He also took part in writing the manuscript and made all the illustrations. M.T. critically read the manuscript and discussed the theoretical work. G.F., G.B. and M.T. made the sample selection and preliminary preparation and characterization. M.S. and R.B. coordinated the work of the beamline personnel J.B., C.K., J.K , F.K., M.R., A.R., T.S., D.Z. who helped to run the beamline, and added his remarks to the manuscript.

**Competing Interests**

The authors declare no competing interests.

**Figures**

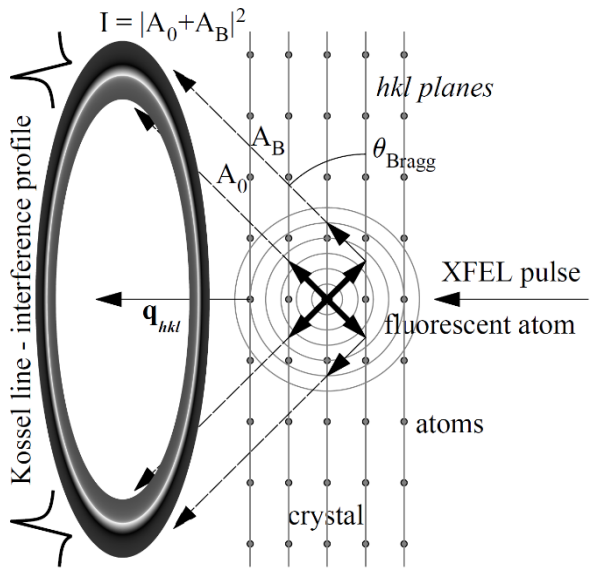

Figure 1. **Formation of Kossel lines.** Interference of x-rays emitted by atoms in a crystal and its Bragg-reflection encodes both the phase and amplitude of the structure factor in the intensity profile of the Kossel line. See also the animation of the illustrated processes in the Supplementary Movie. The figure is adapted from our earlier publicaion[15].

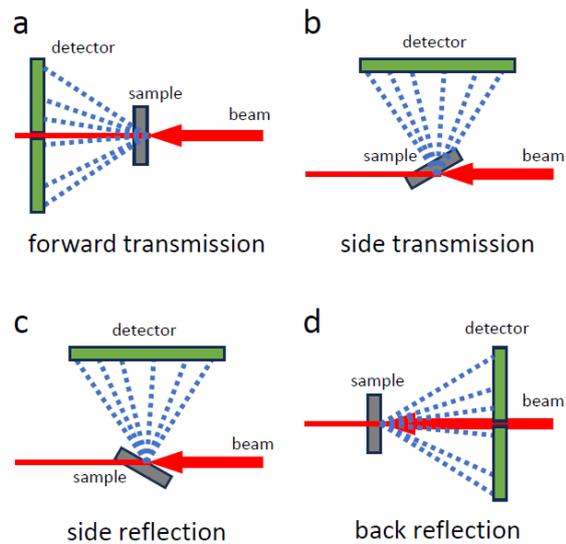
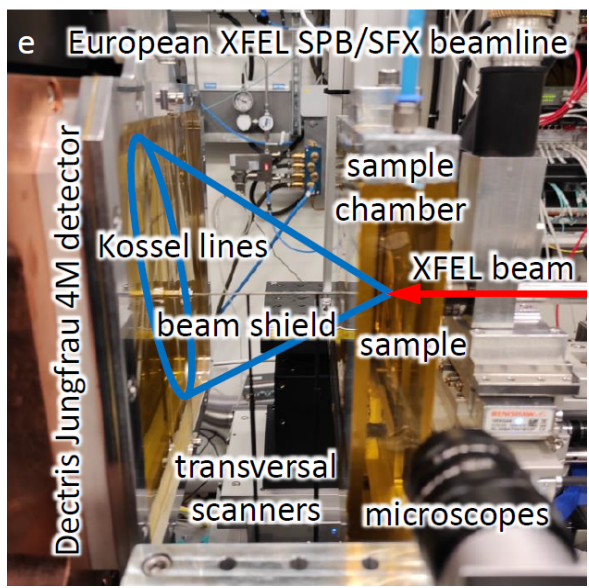

Figure 2. **Experimental arrangement. a,b,c,d** Various transmission and reflection geometries for recording Kossel line patterns from a single XFEL pulse. **e** The experimental setup in forward transmission geometry at the European XFEL.

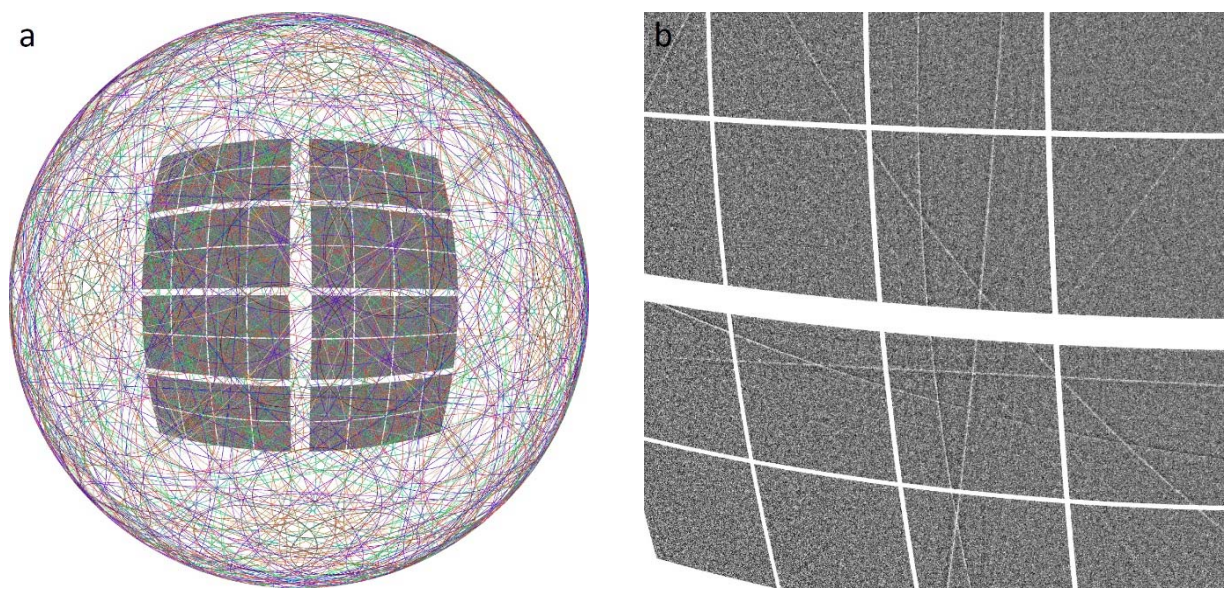

Figure 3. **Experimental Kossel line pattern. a** Complete Kossel line pattern and the observed fraction projected on a sphere. Lines of the same *d*-value have a common color. **b** Magnified region of the background-corrected normalized pattern reveals the stronger lines even to the naked eye. See also the high-resolution Supplementary Images.

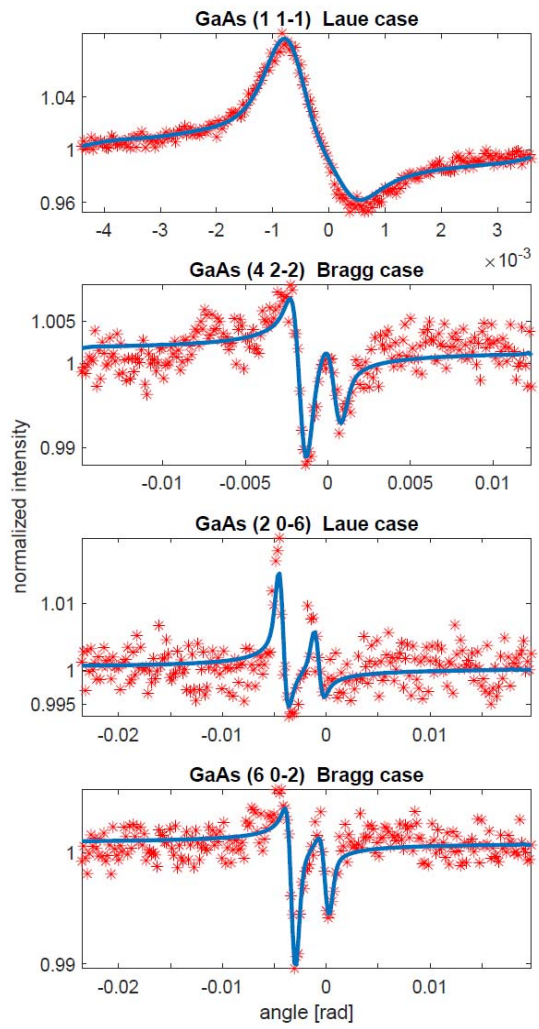

Figure 4. **Selected GaAs Kossel lines.** Line profiles of low (1 1-1), medium (4 2 -2) and high (2 0 -6), (6 0 -2) index reflections for GaAs.

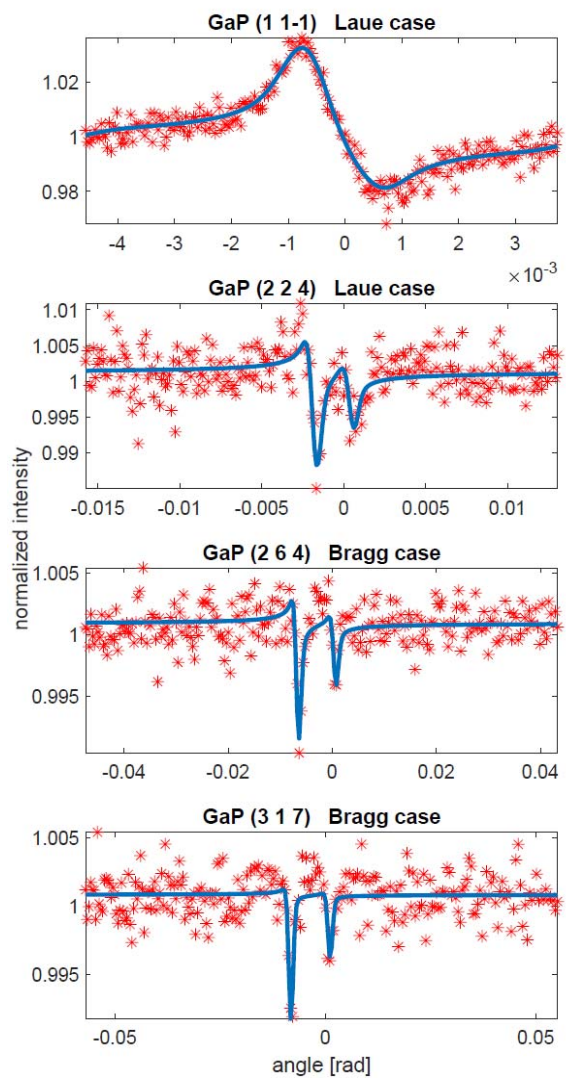

Figure 5. **Selected GaP Kossel lines.** Line profiles of low (1 1-1), medium (2 2 4) and high (2 6 4), (3 1 7) index reflections for GaP.

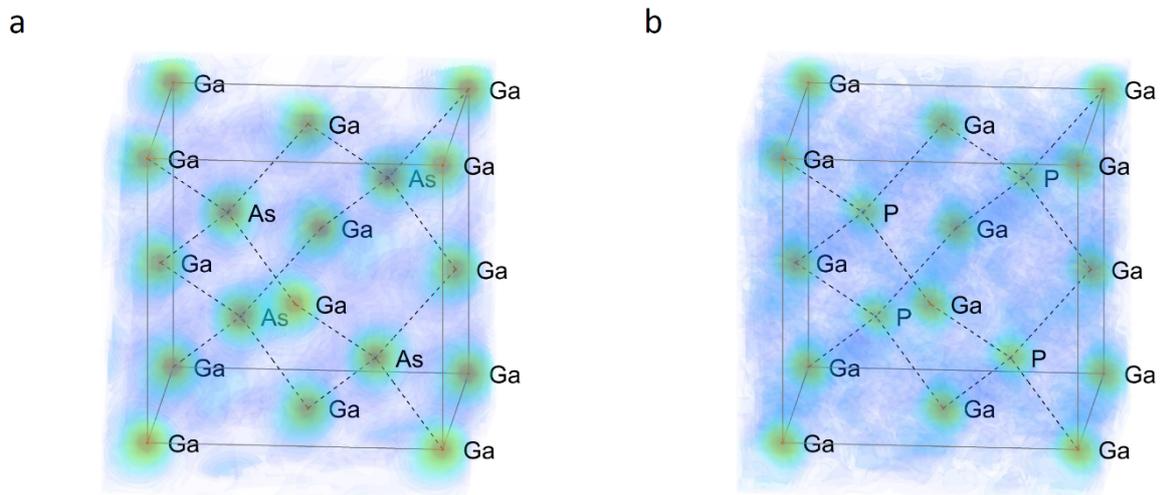

Figure 6. **Reconstructed electron densities. a** GaAs electron density obtained by direct Fourier synthesis from about 100 experimentally obtained structure factor amplitudes and phases. **b** Same for GaP, where the difference between Ga and P electron density is also visible. For 3D rendering see the Supplementary Movies.

Supplementary Information
for
**3D atomic structure from a single XFEL pulse**
by
G. Bortel, M. Tegze, M. Sikorski, R. Bean, J. Bielecki, C. Kim, J. Koliyadu, F. Koua,
M. Ramilli, A. Round, T. Sato, D. Zabelskii, G. Faigel

In the first part of this document we describe the complete evaluation process, from the measured raw image to the 3D real space structure. Parts of this procedure were already described in our previous publications [1,2,3]. This part is followed by a section about the effect of radiation damage and another about the prospective applications.

**Evaluation process from raw Kossel line patterns to electron densities**

*Calibration to photon counts.* The raw detector data represents the total charge measured in each pixel of the 4M pixel Jungfrau detector [4]. This is proportional to both the photon energy and the number of photons detected. Based on the nominal photon energy (fluorescent energy of the excited element of the sample), these pixel values were converted to calibrated photon counts. The calibrated detector images for GaAs and GaP single-shot structure determination are shown in Figure S1. They consist of 8 modules of 8 chips each, with some gaps between them. The calibrated images contain 20/120 and 200/800 fluorescent photons/pixel/pulse at the edges/center of the 4M Jungfrau detector placed at 120 mm from the sample for GaAs and GaP, respectively.

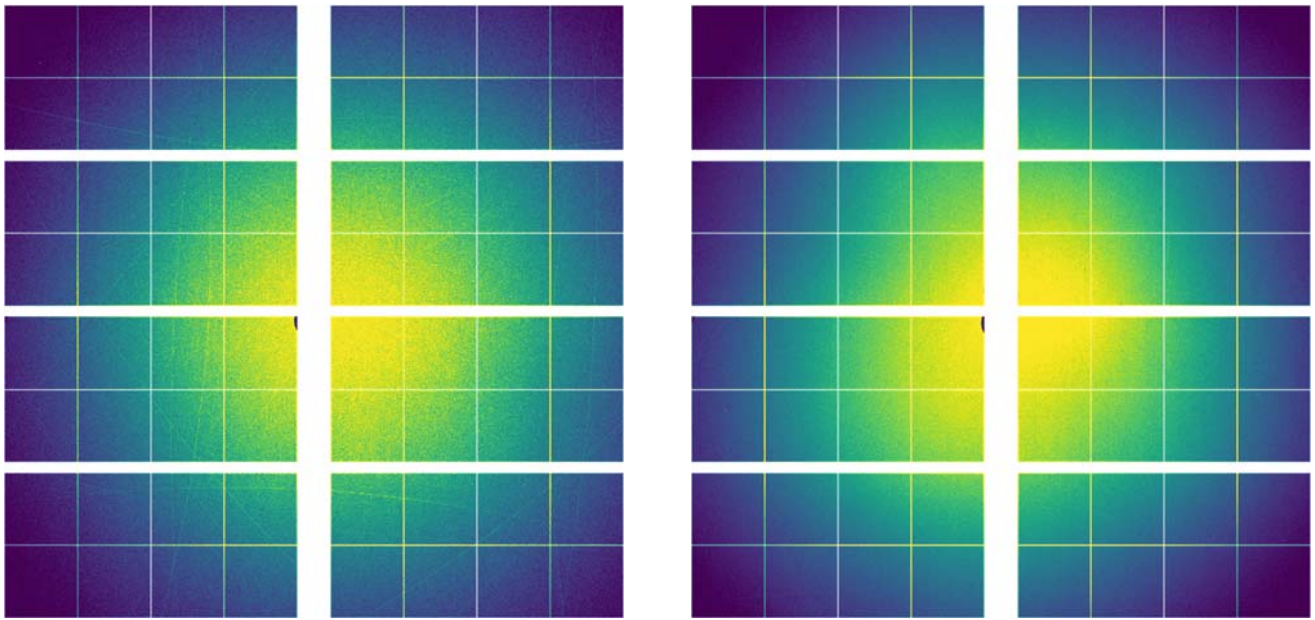

Figure S1. Calibrated detector images for GaAs and GaP. See also full-resolution lossless Supplementary Images. Their size (2178×2170 pixels) exactly correspond to the number of detector pixels. Low and high values of the colormap (photons/pixel) are indicated in the filename.

*Background removal and normalization.* A continuous decrease in intensity from the center to the edge of the detector is caused by three factors: scattering of the incident beam on the gas in the experimental chamber, geometrical effects caused by the flat detector and absorption of the fluorescent beam in the sample and in the gas present in the chamber. This smooth, slowly changing background intensity is a disturbing factor when analyzing the sharp Kossel lines and was removed. The background determined by median filtering was used to normalize the calibrated image. This puts the patterns on a general scale, where 1 corresponds to the level of unmodified fluorescent intensity, and the surplus and deficit intensity in the Kossel line profiles is related to this level. Figure S2 shows these background-corrected and normalized images. The Kossel lines are clearly visible, even to the naked eye.

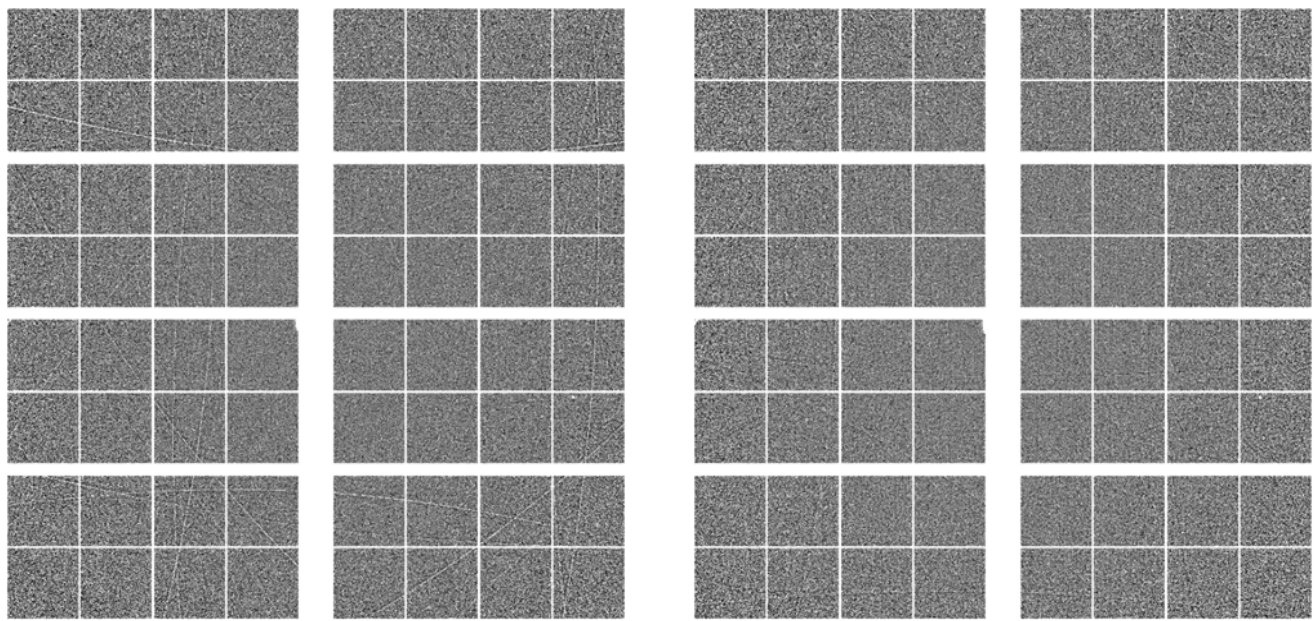

Figure S2. Background corrected and normalized images for GaAs (left) and GaP (right). See also full-resolution lossless Supplementary Images. Their size (2178×2170 pixels) exactly correspond to the number of detector pixels. Low and high values of the grayscale map (normalized units) are indicated in the filename.

*Kossel line indexing.* Our procedure to find both the unknown lattice and orientation of the crystal, *i.e.* autoindexing the pattern was already described earlier [3]. Briefly, conic sections are individually fitted to the lines on the observed image yielding the axis and opening angle of the individual cones. By assuming a common apex for all cones, this information is converted to scattering vectors for each Kossel line, defining a reciprocal lattice point in the reciprocal space. The set of these reciprocal lattice points is indexed with some well-established indexing procedure used in single crystal diffraction yielding the orientation, lattice and reciprocal lattice vector indices.

*Geometry refinement.* A special software tool was developed that allows simultaneous refinement of all relevant parameters: detector module positions, crystal lattice parameters and crystal orientation matrix. The result of this process in a magnified region of the pattern is shown in Figure S3., where color-overlaid lines show the calculated centerline of the $K\alpha_{1,2}$ and $K\beta$ Kossel lines, matching the lines on the image with pixel precision. This refinement also indicates that the lattice is cubic, since a single lattice constant and 90° lattice angles perfectly describe all Kossel lines.

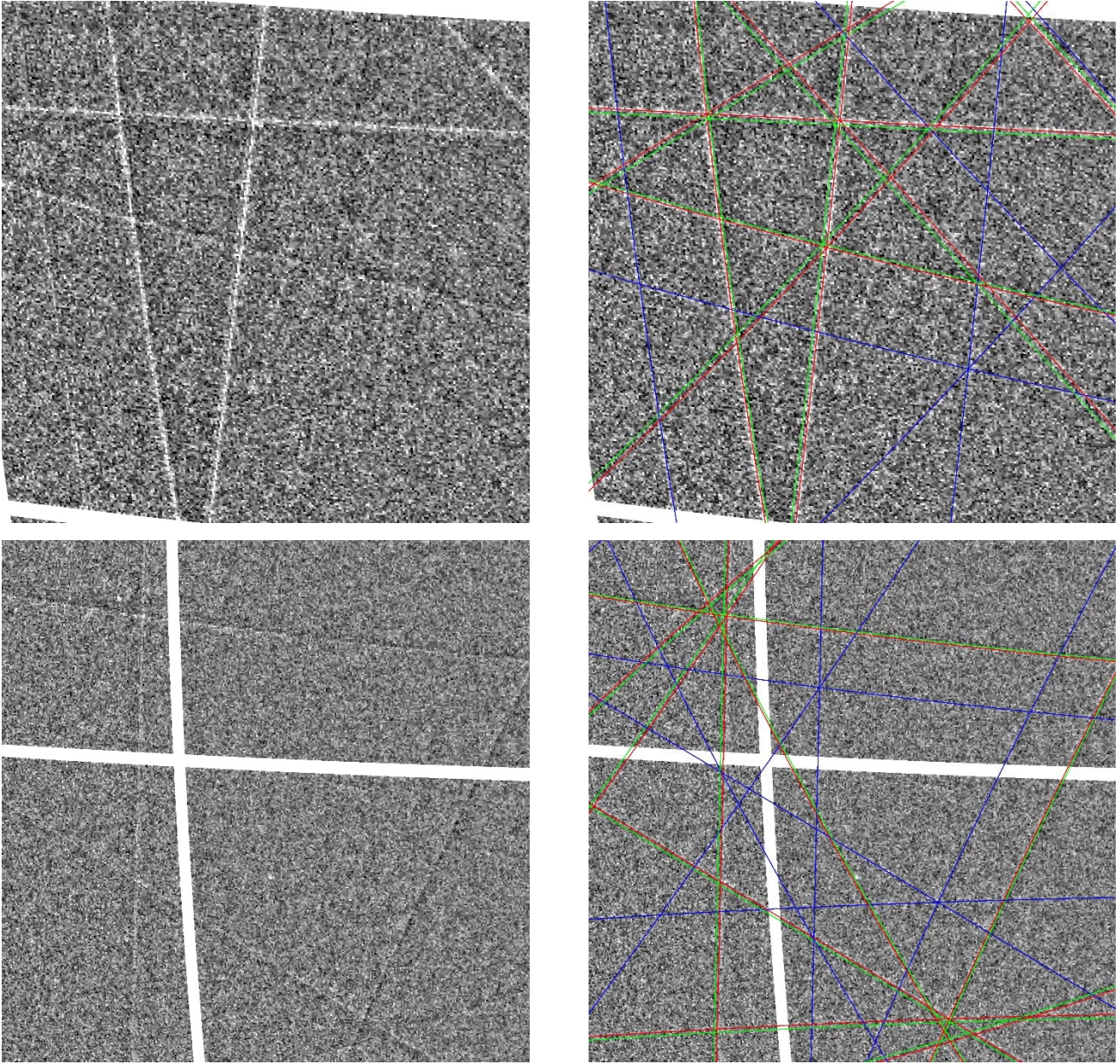

Figure S3. A magnified region of the patterns without and with overlaid indexed lines for all 3 emission energies (red, green and blue) up to 1.4 Å resolution for GaAs and GaP.

*Profile extraction.* The next step in the evaluation process is the extraction of the line profiles. In principle, the line profile is changing along a line *i.e.* around a Kossel cone, since the angle of the involved waves with the sample surface is changing. However, this change is small on the relatively small solid angle covered by the detector. Therefore, we integrated the images along a Kossel line for different cone opening angles to obtain an average line profile. These profiles as a function of the opening angle can be brought to a common scale, the emission wavelength, using Bragg's law. These extracted line profiles together with the number of contributing pixels up to 1.4 Å resolution are shown in Figure S4. The deviation from the average fluorescence level of the strongest Kossel lines was (+20%, –8%) and (+4%, –3%) for GaAs and GaP, respectively. It is worth comparing these amplitudes with estimated Poisson statistics of the extracted profiles: The average ~80 and ~500 photons/pixel raw calibrated intensity (Figure S1) for GaAs and GaP respectively, and the typical 300 contributing detector pixels to a single Kossel line profile point (Figure S4 bottom panels) gives 24000 and 150000 photons/profile point estimated values. Their relative statistical uncertainty, 0.0065 and 0.0026 is well below the line amplitudes, indicating statistically meaningful results.

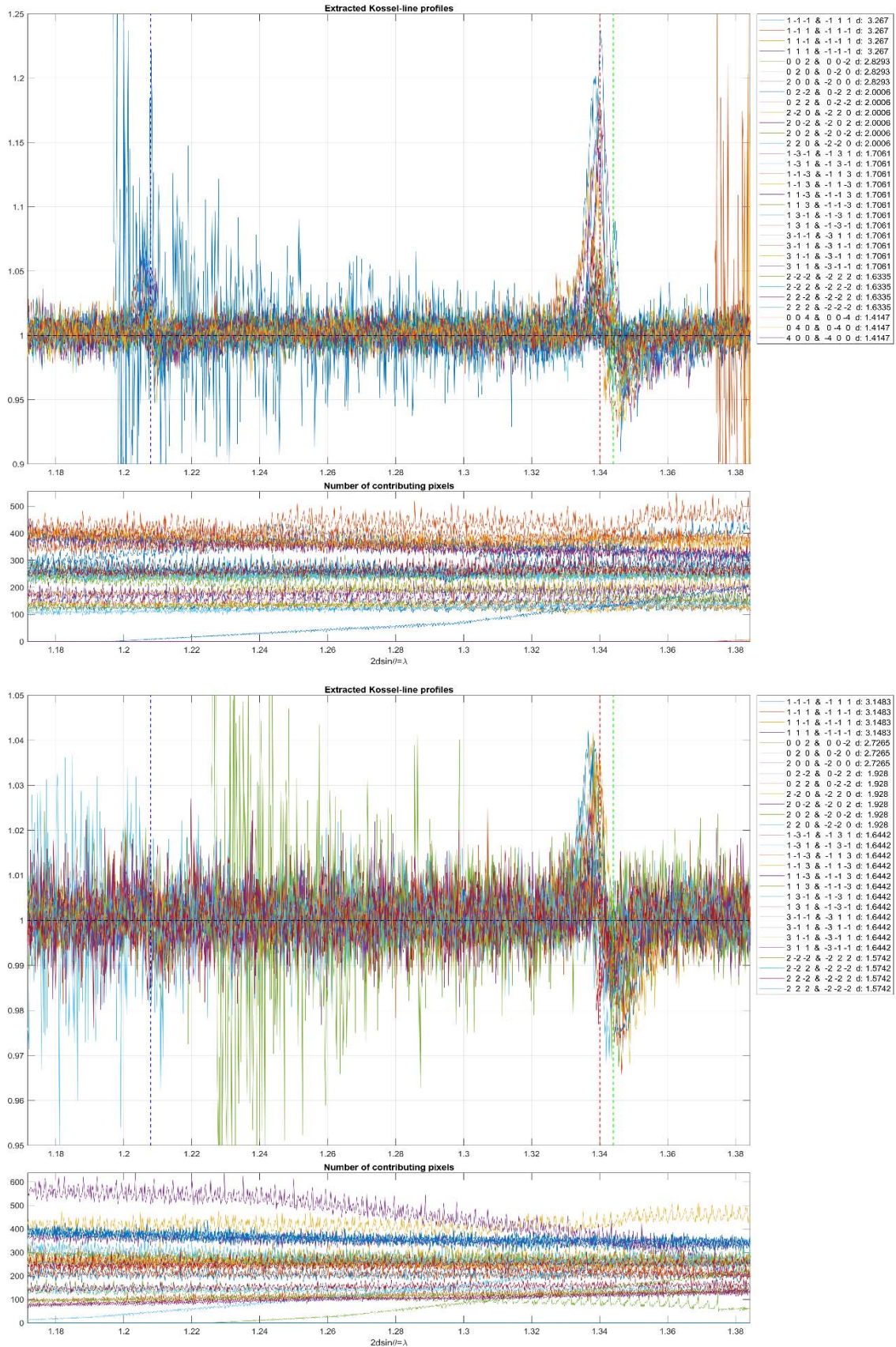

Figure S4. Extracted Kossel line profiles up to 1.4 Å resolution for GaAs and GaP. It can also be seen on the bottom panels, how the statistical noise increases where the integration range is not completely covered by the detector image and the number of contributing pixels decreases to zero.

The inventory of the reflections in the step of profile extraction is the following: The ~0.7Å resolution limit defined by the emission line wavelength and the ~5.55Å average lattice parameter of the structures allow ~2000 cubic reflections within the resolution sphere $[(4\pi/3\,(2\pi/0.7)^3)/(2\pi/5.55)^3]$. In the profile extraction, the Friedel pairs, the opposite halves of a Kossel cone were not distinguished, they contributed to a single profile, reducing the number of reflections to ~1000. The complete absence of Kossel lines in the extracted profiles for all mixed parity indices indicates a volumetric systematic absence, that proves the face centering of the conventional cubic lattice. These reflections were taken as zero, further decreasing the number of usable reflections to ~250. The profiles of all these Kossel lines were attempted to be extracted from the image, but some of the Kossel cones are not covered by the solid angle of the detector at all, some have just a fraction on the detector and only a few high index cones fall completely on the image. The number of contributing pixels to the profile integration (shown also in Figure S4) is the strongest factor that affects the existence, statistics and quality of the extracted profile. A critical selection of these left us with ~100 usable Kossel line profiles for further analysis.

*Profile fitting.* The line profile is used to obtain the phases and the amplitudes of the structure factors. In Figure S5 typical low, medium and high index profiles are shown for illustration (it is the same as Figure 4 and 5 of the paper).

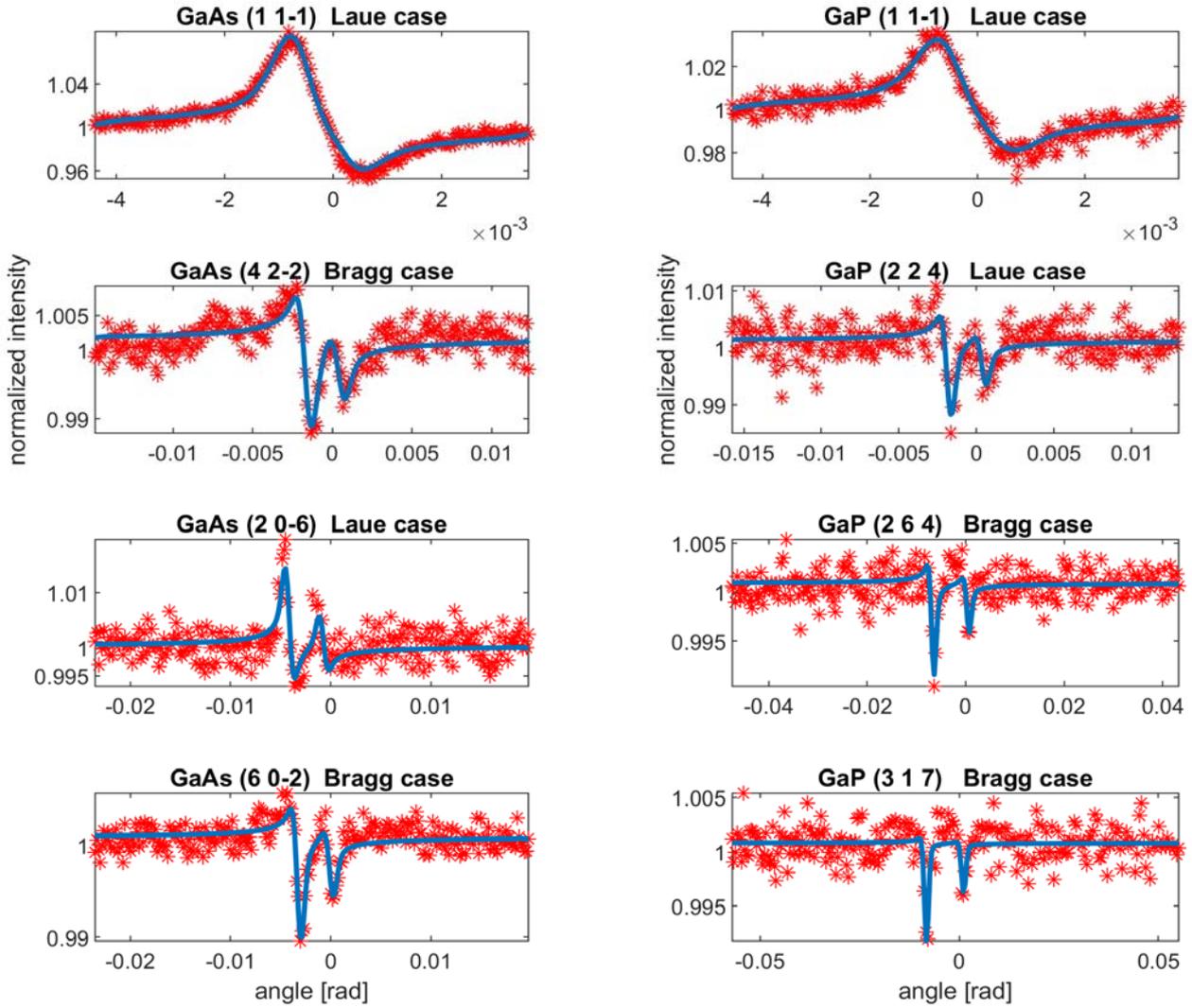

Figure S5. Theoretical fit to the Kossel line profiles for some selected reflections for GaAs and GaP, respectively. Reproduced from the main text to illustrate the complete evaluation process in this document.

For the description of the line profiles, we followed the work of Hannon and Trammell [5,6,7]. In their theoretical treatment, they used the real space approach, which starts from the Darwin description of dynamical diffraction of x-rays in single crystals. However, they extended the theory for the case of inside sources. The theoretical treatment is quite involved, and we do not repeat it here. We give only the final formulas used in the evaluation. Our notation mostly follows the convention of [5,6,7] for easier understanding. The solution to this diffraction problem splits into two parts, depending on the geometry of the experiment: (i) Laue case and (ii) Bragg case. In the terminology of dynamical diffraction, these cases are explained with the help of the crystal surface, the crystallographic layers and the beam directions. The two cases are illustrated in Figure S6.

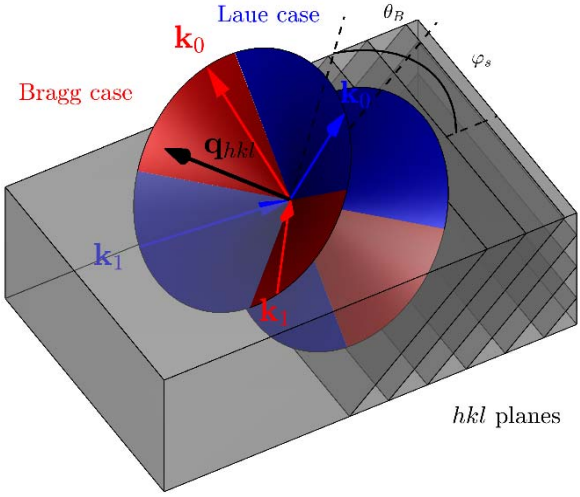

Figure S6. Illustration of Bragg and Laue cases. In the Bragg case, the wave vectors coupled by the reflection are on one side of the crystal surface, while in the Laue case are on opposite sides. To see how the Bragg and Laue cases change with the surface angle, see the Supplementary Movie. The figure and movie are adapted from our earlier publication [1].

The variation of the intensity as a function of the deviation from the Bragg angle is given by (1) and (2) for the Laue and Bragg cases, respectively. These formulas express the contribution of a source atom located at $r_i$ in the $M_j$-th layer to the reflection $hkl$, at a given wavelength $\lambda$.

For the Laue case:

$$I(\delta) = B \left| \left(1 + \frac{\nu}{\beta} - C\frac{F^{01}}{\beta}\right) e^{iM_j(\alpha+\beta)} + \left(1 - \frac{\nu}{\beta} + C\frac{F^{01}}{\beta}\right) e^{iM_j(\alpha-\beta)} \right|^2, \qquad (1)$$

where

$\alpha = \frac{1}{2}(F^{00} + F^{11})$,
$\beta = (\nu^2 + F^{01}F^{10})^{1/2}$,
$\nu = \delta + \frac{1}{2}(F^{00} - F^{11})$,
$C = e^{i(\mathbf{k}_0 - \mathbf{k}_1)\mathbf{r}_i} e^{-i(\alpha+\delta)}$.

For the Bragg case:

$$I(\delta) = B \left| \frac{1 + im^* C F^{01}}{Q} e^{iM_j(\alpha+\beta+gd-\delta)} \right|^2, \qquad (2)$$

where

$\alpha = \frac{1}{2}(F^{00} - F^{11})$,
$\beta = (\nu^2 - F^{01}F^{10})^{1/2}$,
$\nu = \delta + \frac{1}{2}(F^{00} + F^{11})$,
$m^* = \dfrac{i\left(1 - e^{2i(M - M_j + 1)\beta}\right)}{\nu + \beta - (\nu - \beta)e^{2i(M - M_j + 1)\beta}}$,
$Q = \dfrac{\nu+\beta}{2\beta}\left(1 - \dfrac{\nu-\beta}{\nu+\beta} e^{2i(M - M_j + 1)\beta} + i(M - M_j + 1)(\nu - \beta)\left(1 - e^{2iM_j\beta}\right)\right)$,

$C = e^{i(\mathbf{k}_0 - \mathbf{k}_1)\mathbf{r}_i}$.

$\delta$ is related to the angular deviation from the kinematical Bragg angle $\theta_B$; $M$ is the total number of layers, $M_j$ is the serial number of the layer containing the source atom below the surface of the crystal toward the detector, , $\mathbf{k}_0$ is the direct wave wave-vector, $\mathbf{k}_1$ denotes the wave-vector of the wave which is scattered into $\mathbf{k}_0$ satisfying the Bragg condition ($\mathbf{k}_0 = \mathbf{k}_1 + \mathbf{q}_{hkl}$ where $\mathbf{q}_{hkl}$ is the reciprocal lattice vector of the reflection under consideration and $|\mathbf{k}_0| = |\mathbf{k}_1| = 2\pi/\lambda$), $g$ is the projection of $\mathbf{k}_0$ to the normal of the crystal surface, $\mathbf{r}_j$ denotes the position of the source atom relative to the origin of the unit cell, $B$ is a normalization factor, which describes the intensity of the Kossel line relative to the total off-Bragg intensity and can be calculated from the fluorescent yields. $\theta$ is the rocking angle, $\varphi_1$ ($\varphi_0$) is the angle between $\mathbf{k}_1$ ($\mathbf{k}_0$) and the crystal surface, $d$ stands for the interplanar spacing, $F^{00}$, $F^{11}$, $F^{01}$ and $F^{10}$ are the planar scattering amplitudes (for definition see equations (7) and (8) in [6]).

The total intensity variation is obtained by summing the contributions of all layers containing source atoms. Since we intend to determine the phase ($\Phi$) and magnitude ($A_0$) of the structure factor we recall the dependence of $F^{01}$ on $\Phi$ and $A_0$ [6]: $F^{01} = A_0/\sin(\varphi_1)e^{i\Phi}$. Note, that in our measurement, the source atom fixes the origin and through this the phases of the Kossel lines, unlike in traditional crystallography, where the choice of the origin of the unit cell affects the phase of reflections via a phase-factor. In our measurement the source atom fixes the origin and through this the phases of the Kossel lines. Profiles were fitted to the lines according to equations (1) and (2). In principle, there are four parameters to fit: $F^{01}$, $F^{10}$, $F^{00}$ and $F^{11}$. Examining equations (1) and (2) we see that $F^{00}$ and $F^{11}$ forward scattering amplitudes determine the shift of the diffraction lines relative to the kinematical Bragg angle. Their absolute value is proportional to the number of electrons in the unit cell, and their imaginary part is determined by the absorption. These values are known and given by the sample composition. In general, their values are small; the shifts are in the arcsec range. Since we cannot measure absolute angular positions with this precision by our setup, we fit the position of the peaks for every line independently. However, the fitted values of the peak positions are not used in the structure solution. For the structure factor determination, the important parameters are $F^{01}$ and $F^{10}$. Their phase determines the shape of the Kossel lines. By definition, $F^{01}$ and $F^{10}$ are not independent; $F^{01} = A_0/\sin(\varphi_1) \times e^{i\Phi}$ while $F^{10} = A_0/\sin(\varphi_0) \times e^{-i\Phi}$. This means that we have to fit the Kossel line profile by a minimum of 3 parameters: $A_0$, $\Phi$ and a position parameter. However, in practice we have one more fitting parameter determined by the experimental conditions: the line broadening. It comes from three factors: the energy width of the fluorescent lines, the imperfection of the crystal, and the angular resolution of the experimental setup. The energy width of the fluorescent lines can be precisely given but the other two factors change from line to line depending on the geometry and, in practice, they cannot be easily derived. Estimating their contributions, we found that the crystal imperfection and the angular resolution of the setup dominate, and the smallest contribution comes from the energy width of the fluorescent lines. We took into account the line broadening by convoluting the theoretical lines with a Gaussian. An average value for the $\varphi_1$ and $\varphi_0$ angles of the wavevectors with the sample surface (more precisely, the ratio of their sines) were taken from the indexing and geometry refinement step for the fraction of the Kossel line falling on the detector.

Therefore, all together the line profiles were fitted by four parameters: line position, line broadening, amplitude of structure factor and phase of the structure factor. Since the statistics of the measurements were relatively poor because of the single pulse used for the collection of a full pattern, the fitting procedure had to be done very carefully. A brute force approach *i.e.* starting from a random set of all parameters and fitting all in one iteration process does not converge. Therefore, we used a 3-stage process: first, we found approximate starting parameters for the iteration. This was done by choosing estimated values for the position, broadening and magnitude of the lines, meanwhile mapping the phase ($\Phi$) from 0–360° in 20° steps. We selected the starting phase at the value where the line shape was closest to the measured one. This was done for all lines by visual inspection. In the second stage, we fixed the phase and performed a 3-parameter nonlinear minimization of the sum of squared differences of the measured and calculated profiles. This gave the experimental parameters: the width of the Gaussian, and the positions of the lines, and a starting value for the amplitudes to the last iteration. In the third stage, we fixed the experimental parameters and iterated for the phase and amplitude.

The result of the above fitting procedure for the two experimental parameters is the following: the broadening is between 30-80 times, while the line shift is -10 to +20 times the theoretical line width. These are in accordance with the expectations based on the detector geometry. As we mentioned earlier, we did not use these parameters in the structure solution. The two important parameters concerning the structure solution are the phase and magnitude of the structure factor. Most of the phases of the structure factors were within 20 degrees of the theoretical values. However, some phases had as large as 40 degrees errors. We show in Figure S7 the measured phases against the theoretical values. The measured amplitudes of the structure factors are shown in Figure S8 together with the theoretical values.

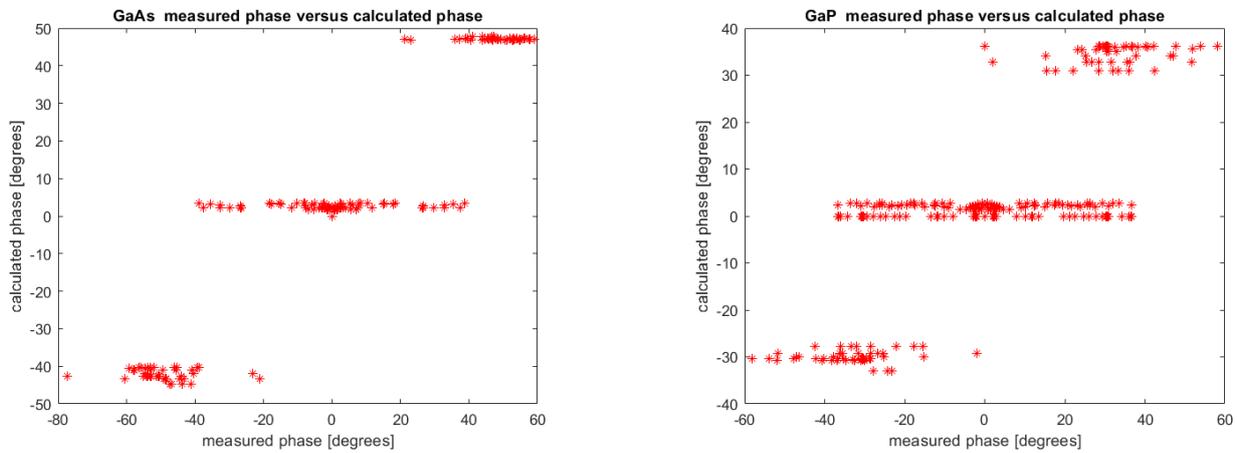

Figure S7. Correlation Figures of the measured and calculated structure factor phase angles for GaAs and GaP.

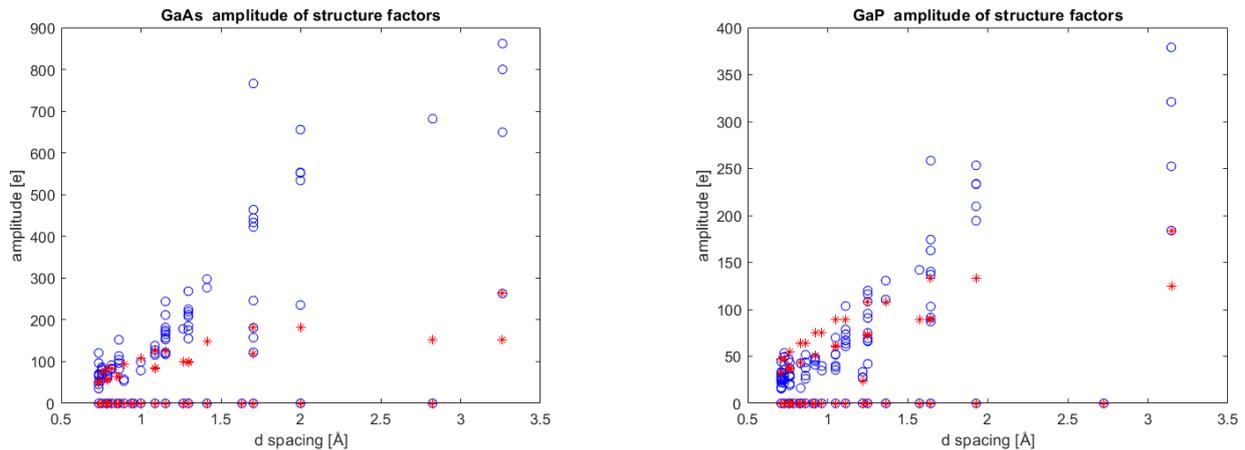

Figure S8. Measured (blue circles) and calculated (red stars) structure factor amplitudes for GaAs and GaP.

*Fourier-synthesis.* The last step of evaluation is the inversion of the structure factors to real space electron density, the atomic structure. We did this by putting the complex structure factors to the measured reciprocal lattice points, extended the reciprocal lattice with the Friedel pairs (equivalent of supposing real electron density) and for the forward direction, which we could not measure we put the total charge in the unit cell given by the composition of the sample. For smoothness, we extended the Fourier transformation range by three times the size of the range of the measured indices. The unknown lattice points were taken as zeros. This 3D complex matrix was Fourier transformed, resulting in a real electron density distribution, shown in Figure S9.

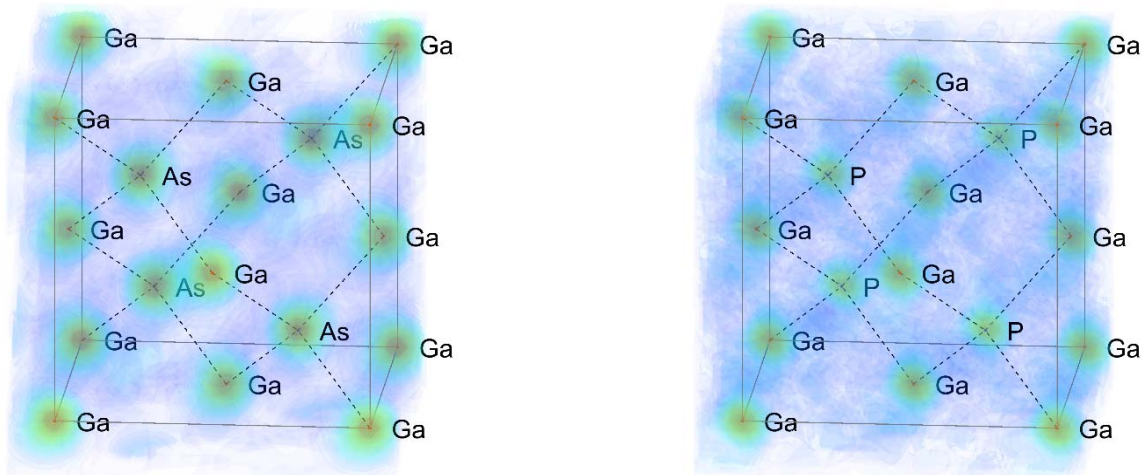

Figure S9. Electron density distribution obtained by direct Fourier synthesis for GaAs and GaP. For 3D information see the Supplementary Movies.

In order to estimate the error in atomic positions and electron density (peak heights and width where atoms are found), we made model calculations on the same grid as the measured data were given using the same number and *hkl* index of structure factors that were measured. We compared this to the density reconstructed from the measured data. In Figure S10 we show the reconstructed electron density of GaP for the theoretical structure factors and for the measured structure factors using isosurface plot. On this type of plot, it is easier to recognize the differences between the two reconstructions. It is clear that the "size" of the atoms are larger in the reconstruction from the measurement. This is caused by the error in the measured structure factors. The difference in width is about 15%. However, the positions of the peaks do not change within the grid size, which is determined by the FFT range. Further, we also checked the integrated intensity within the peaks, and we found that the deviation compared to the theoretical values was in the range of 10%. We made the same calculations for GaAs and found similar results. This explains why we could not distinguish the Ga and As atoms in our measurements; the error in the integrated intensity was larger than the difference in the atomic numbers.

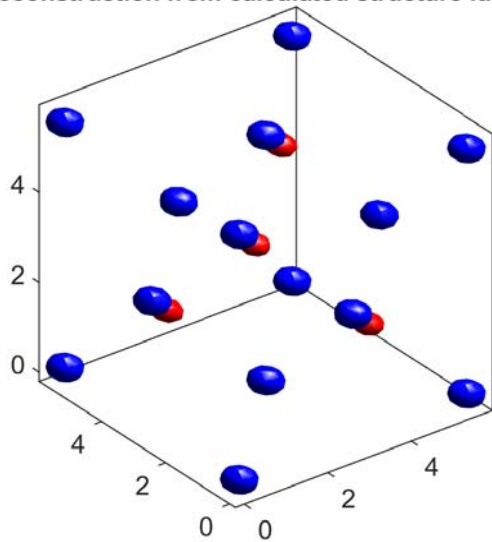 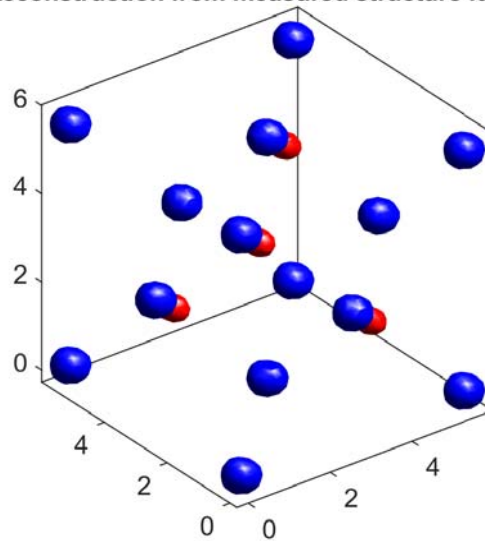

Figure S10. Comparison of real part of reconstructed electron densities with calculated and measured structure factors using the same grid and set of reflections. In both figures the isosurface is drawn at 50% level of peak density.

## Discussion of radiation damage

In XFEL experiments, the extremely high number of photons interacting with the sample in a small area and in a very short time causes radiation damage. This, in most of the cases is fatal, resulting in evaporation of the sample (Figure S11). However, there is a possibility to overcome this problem by measuring the structure before the atoms have time to move. This was suggested first in 2000 [8]. Later, many theoretical works investigated radiation damage (see for example [9]), and the idea was also experimentally proved. In our case, the most relevant experiments are the serial femtosecond crystallography (SFX) experiments, where small crystallites are introduced into the beam. Using SFX, several structures were determined at atomic resolution. There were studies aiming to find the effect of pulse length and pulse intensity on the structure solution using SFX. In an early work [10] it was found that in the 1 μm spot size range, at about 10 fs pulse length and at $\sim 10^{12}$ photons/pulse region the effect of radiation damage resulted in higher background, but the diffraction spots could be measured, and the structure could be determined without problem. In our experiment the spot size was 25 μm, which is about 10 times the spot size used in SFX measurements. The other beam parameters are the same as in an average SFX experiment. Therefore, the deposited energy on unit area is 100 times smaller than in a typical SFX. Since our measurement is also x-ray diffraction (although using inside sources), we do not expect deformation in the diffracted intensity pattern. Further, even the increase of the background is negligible because of the 100 times less deposited energy in unit area. Besides the radiation damage nonlinear effects could also appear, caused by the high energy density in the sample. This type of effects was studied in [11]. In this work the authors calculate the effect of high energy density on the imaginary part of the atomic scattering factor f'. They find significant changes for samples used in SFX experiments. However, we can neglect this effect because we have about 100 times smaller energy density on our sample. Although, after the pulse we see a hole where the beam hits the sample (Figure S11), the formation of this hole happens in a much longer time scale than the measurement of the elastically scattered fluorescent radiation. However, the radiation damage prevents the measurement of the original structure on the same spot of the sample using a second pulse.

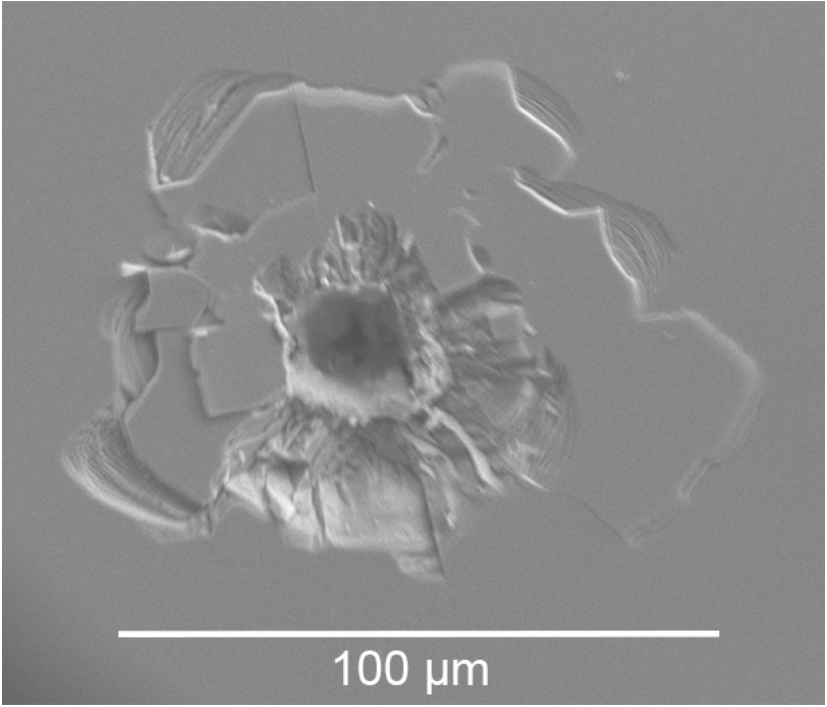

Figure S11. Scanning electron microscope image of the GaAs sample after hit by a single XFEL pulse. The hole in the center reflects the ~25μm diameter of the beam. Around the hole we see the deterioration of the crystal surface caused by the heat shock.

**Prospects for possible applications**

In this section we summarize the conditions necessary for measuring Kossel line patterns at XFEL-s. These conditions give guidelines for those who intend to use our method for structural studies. Based on the conditions in the last part of this section we give few specific examples for application. First, we discuss the sample parameters and then the beam parameters.

The sample has to contain at least one heavy element in the unit cell. Heavy element means emission energy larger than 5 keV. Smaller energy will result in lower spatial resolution. The sample size has to be larger than 10 μm. At smaller sample size one has to use smaller focal spot (see beam parameters later in the text), which may result in higher background and nonlinear effects modifying the atomic scattering factors leading to distortion in the diffracted intensity distribution. This would increase the error in the phase determination from the Kossel line profile. In our demonstration experiment, we used samples with simple structures, only a few atoms in the unit cell. In the next part, we will discuss prospects for more complicated structures.

First, take a sample containing one heavy atom and many other atoms, which are not excited. In this case, the structure solution is going the same way as we did in the presented two cases. However, as the number of atoms increases in the unit cell one has to measure more Kossel lines. As a rule of thumb, one should measure about 10 Kossel lines for each atom in the unit cell.

Our next example is when we have a sample containing two different heavy atoms, both of which are excited. Although in principle it is possible to experimentally separate the emission lines if their energies are far enough, however we would not rely on this. Instead, we should sort out the Kossel lines corresponding to the two emission energies by their different positions in the measured 2D image. This was already demonstrated in our earlier experiment performed at synchrotron on GaAs [1], where both the Ga and As emission lines were used. If the lines are sorted out this way, the solution of the structure is even more stable, because it works as if two independent measurements with different energies were done.

The third example is a sample in which there are two inequivalent sites of the same atom type in the unit cell and many other atoms, which are not excited. In this case, the phase of the structure factors cannot be determined without additional knowledge. We have to know the relative position of the two excited atoms. In principle, this could be determined fitting the relative position together with the phases for all lines in one iteration process. However, this would make the solution very complicated. Therefore, we do not recommend to measure this type of samples before working out the proper procedures for the above mentioned iteration process.

One more remark about the sample composition: so far, we mostly discussed what type of fluorescent atoms the sample has. However, the measured pattern contains information on the non-fluorescent atoms too. The question arises: what we can tell about those. Similarly to traditional single crystal diffraction measurements, the lighter elements have smaller weights. Therefore we can measure on samples containing light elements (for example biological samples), but we do not expect to see lighter elements than carbon. Further, as we mentioned in connection to the first example, we have to measure more Kossel lines for more complicated structures, which makes the evaluation more difficult. The reason is that Kossel lines are not point like (as Bragg peaks) but lines on the detector surface. Therefore, it will be more difficult to separate them if there are too many. This results in more stringent technical limits on the detector and evaluation software than we have in traditional crystallography. A further point, which one has to take into account when measuring biological samples, is the crystal quality, which is usually not too good. Mosaic crystals would wash out the line shape resulting in the loss of phase information. However, this does not limit structure solution more than the missing phase in traditional diffraction. In this case, one can use the well-established single crystal algorithms to solve the structure.

Turning to the beam parameters, it has to be tailored to the given experiment. First, the beam energy should be chosen to excite the heavy atom we want to use for producing the Kossel lines. It is advantageous if we excite only a single atom type. The beam size should be about the same as the sample size. It is important to excite the largest number of atoms possible. The pulse should contain as many photons as we can get. Going below $10^{12}$ photons/pulse will severely limit the complexity of the structure we can determine. The length of the pulse should be in the 10-50 fs range to avoid deterioration of the Kossel lines because of radiation damage.

Below we will indicate what type of studies would mostly benefit from single pulse structure determination. We have already mentioned in the main text those general areas, the extremely non-ambient condition studies, which would benefit most. Here we give two specific examples.

(a) Many studies aim to find the atomic structure of matter at very high pressures (for example pressures present at the interior of planets like Earth, Jupiter etc.), which is difficult to statically maintain. In these cases, we cannot do pump-probe experiments in easily repeatable and well-defined way. However, we can make this large pressure for a very short time (ms-ns ). There is no method, which could determine the atomic structure at these cases from one single measurement. Our approach could do it.

(b) Other example is very high magnetic fields (above 30 Tesla). If one intends to study matter in very high magnetic fields, which cannot be produced as a static field, but can be produced as one very short pulse, again there is no method which could determine the 3D atomic arrangement induced by this field. These types of investigations include superconductivity, strongly correlated and quantum critical systems, frustrated quantum magnets etc.

At last, we would like to mention one more possible application area, time resolved pump-probe experiments at XFEL. Presently this type of experiments are done in SFX mode. In this case one has to measure at every time point tens of thousands of samples. This takes long beam time. If the samples, which we intend to study satisfy the conditions for Kossel measurement, we have to take only single shots at every time point (or a few shots taking into account the stochastic nature of the XFEL pulses and the less than 100% hit rate). This would shorten measuring times significantly. Therefore, it would increase the output of these very expensive facilities.